\documentclass[12pt]{iopart}

\usepackage{iopams}
\usepackage[colorlinks=true,linkcolor=blue,citecolor=green,urlcolor= magenta]{hyperref}

\def\bz{\bar z}
\def\bs{\bar s}
\def\p{\partial}
\def\bp{\bar\partial}
\def\ba{\bar a}
\def\bb{\bar b}

\def\l{\lambda}

\newcommand{\szego}{Szeg\"o\ }

\renewcommand{\Re}{{{\rm Re}\,}}
\renewcommand{\Im}{{{\rm Im}\,}}

\renewcommand{\epsilon}{\varepsilon}

\newcommand{\kahler}{K\"ahler }

\newcommand{\R}{{\mathbb R}}
\newcommand{\C}{{\mathbb C}}

\newcommand{\kcalomega}{\mathcal{K}_{[\omega_0]}}

\newcommand{\dbar}{\bar\partial}
\newcommand{\ddbar}{\partial\dbar}

\newcommand{\E}{{\mathbf E}}

\newcommand{\vol}{{\rm Vol}}

\newcommand{\Bcal}{\mathcal{B}}

\newcommand{\hcal}{\mathcal{H}}

\newcommand{\kcal}{\mathcal{K}}

\newcommand{\ocal}{\mathcal{O}}

\begin{document}

\title[Simple matrix models for random Bergman metrics]{Simple matrix models for random Bergman metrics}

\author{Frank Ferrari$^1$, Semyon Klevtsov$^{1,2}$ and Steve Zelditch$^{3}$}

\address{$^{1}$Service de Physique Th\'eorique et Math\'ematique, Universit\'e Libre de Bruxelles\\
and International Solvay Institutes, Campus Plaine, CP231, 1050 Bruxelles, Belgium}
\address{$^{2}$ITEP, B. Cheremushkinskaya 25, Moscow 117218, Russia}
\address{$^{3}$Department of Mathematics, Northwestern  University, Evanston, IL 60208, USA}

\eads{\href{mailto:frank.ferrari@ulb.ac.be}{frank.ferrari@ulb.ac.be}, \href{mailto: semyon.klevtsov@ulb.ac.be}{semyon.klevtsov@ulb.ac.be}, \href{mailto:s-zelditch@northwestern.edu}{s-zelditch@northwestern.edu}}

\begin{abstract}
Recently, the authors have proposed a new approach to the theory of random metrics, making an explicit link between probability measures on the space of metrics on a K\"ahler manifold and random matrix models. We consider simple examples of such models and study the one and two-point functions of the metric. These geometric correlation functions correspond to new interesting types of matrix model correlators. We provide in particular a detailed study of the Wishart model, where we determine the correlation functions explicitly. We find that the random measure in this model turns out to be concentrated on the background metric in the large $N$ limit.

\end{abstract}

\section{Introduction}

This paper is a continuation of a series of papers by the authors on random \kahler metrics. In Refs. \cite{FKZ0,FKZ} we have proposed a new method to define a random metric on a \kahler manifold $M$. The purpose of this paper is to give a detailed study of simple models of random metrics where our methods apply. Our approach is based on the approximation of the \kahler metrics in a fixed K\"ahler class by the Bergman metrics. The latter are parameterized by positive-hermitian matrices $P_{ij}$ as follows
\begin{equation}
\label{bm}
g_{a\bb}(z)=\frac1k\p_a\bp_{\bb}\log \bs_i(z)P_{ij}s_j(z).
\end{equation}
Here the matrix $P_{ij}$ is contracted with a vector $s_i(z)$ (and its complex-conjugate $\bs_i(z)$) defining a basis of holomorphic polynomials of degree $k$ on the manifold, or more precisely, a basis of global holomorphic sections of some line bundle $L^k$ on $M$. The dimension $N_k$ of the vector depends on $k$ and on the complex dimension $n$ of the manifold. At large $k$, we have $N_k\sim k^n+\mathcal O(k^{n-1})$. The positivity of the metric (\ref{bm}) is ensured by the condition that the matrix $P$ is hermitian positive definite. Note that any such $P$ can be written as $P=A^\dagger A$ for some $A\in GL(N_k,\mathbb C)$ defined up to multiplication by a unitary matrix on the left. Note also that constant rescalings of $P$ do not change the metric (\ref{bm}). Thus the space of Bergman metrics $\Bcal_k$ is equivalent to the symmetric space
\begin{equation}
\Bcal_k=SL(N_k, \mathbb C)/SU(N_k)\, .
\end{equation}

The key fact \cite{Ti,Z,C} is that in the $k\rightarrow\infty$ limit, the space $\Bcal_k$ of Bergman metrics of degree k tends
in a strong metric sense to the infinite
dimensional space of all \kahler metrics. As a consequence, many geometric properties of the \kahler metrics can be studied using the finite-dimensional spaces $\Bcal_k$, see e.g. \cite{PS2} for review. With this correspondence in mind, we define a random metric on the \kahler manifold $M$ as the large $k$ limit of a random Bergman metric.
Equivalently, formula (\ref{bm}) shows that defining a random Bergman metric is the same as defining a random positive definite hermitian matrix $P$. 

Probably the best known models of random metrics in physics are Polyakov's models of two dimensional CFTs coupled to gravity based on the Liouville action \cite{P}, see \cite{DS} for recent rigorous treatment. Other geometric actions, like the Mabuchi functional, were recently shown to play a r\^ole when non-conformal matter is coupled to gravity \cite{FKZ0,FKZ1}.

Our goal in this paper is rather to study a much easier class of random measures on the space of metrics, which are natural from the point of view of the parametrization (\ref{bm}) of the metrics in terms of the matrices $P$. Namely, we consider the simple class of matrix models for which the random matrix measures $d\mu(P)$ depend only on the eigenvalues of $P$. In other words, the measures we consider are invariant under conjugation of $P$ by a unitary matrix $U\in U(N_k)$,
\begin{equation}
\label{uninv}
d\mu(U^\dagger PU)=d\mu(P)\, .
\end{equation}
A particular example that we shall study in full details is the so-called Wishart model for which
\begin{equation}
\label{Wis}
d\mu(P)= e^{-g\,\tr P} [dP]_{\rm Haar}\, ,
\end{equation}
where $[dP]_{\rm Haar}$ is the standard Haar measure on positive hermitian matrices defined in Section \ref{secMM} and $g>0$ is a parameter.

Matrix models of the type (\ref{uninv}), and in particular (\ref{Wis}), are standard matrix models whose large $N$ limits have been extensively studied using well-known matrix model technology. However, the geometric interpretation in terms of random metrics given by the formula (\ref{bm}) suggests to study an entirely new class of correlators
involving the products of the logarithms
\begin{equation}
\label{integrals}
\int  \log \bs_i(z_1)P_{ij}s_j(z_1)\cdots\log \bs_i(z_m)P_{ij}s_j(z_m)\, d\mu(P)
\end{equation}
depending on the points $z_1,\ldots,z_m$ on the manifold. These correlators are nontrivial because they do not depend only on the eigenvalues of the matrix $P$. If we decompose $P=U^{\dagger}\Lambda U$, where $U$ is a unitary matrix and $\Lambda$ is diagonal, we thus have to perform a nontrivial integral over $U$. This can be done using the
Harish-Chandra-Itzykson-Zuber formula \cite{HC,IZ}. The result depends on the coordinates $z_p$ on the manifold via various inner products between the vectors of sections $s_i(z_p)$. These inner products turn out to be proportional to the Bergman kernel. We are then able to find the large $k$ (equivalently, large $N_k$) limit of the correlators using the well-studied asymptotic expansions for this kernel \cite{MM,LS}.

Our main results include explicit computations of one and two-point functions (\ref{integrals}) for general measures of the type (\ref{uninv}) for finite values of $k$, see Eqs. (\ref{corr2}, \ref{final2p}) in the main text. The one-point functions turn out to be very simple and universal, independently of the precise form of the measure, see Eq. (\ref{corr3}). The finite $k$ two-point functions depend on the details of the measure, but their large $k$ limit is universal and depend only on a model-dependent constant, see Eq. (\ref{res}). In the case of the Wishart model (\ref{Wis}), we go further and find an explicit formula Eq. (\ref{answer}) for the two-point function at finite $k$. The main feature of the measures considered here is that in the large $k$ limit they turn out to be heavily concentrated around the background metric. In particular, the results for the Wishart model are consistent with the delta-function like distribution of the space of all K\"ahler metrics. This feature is an artefact of using the matrix $P$ as a basic random variable. As was pointed out in \cite{FKZ}, we expect much smoother limiting measures when using the ``tangent space'' random variable $\frac1 k\log P$ instead of $P$. Nevertheless, we believe that the present work provides a necessary basis for further investigations of the random measures on $\Bcal_k$.

The paper is organized as follows. In Section \ref{background} we review basic facts on Bergman metrics and collect some formulas on the expansion of the Bergman kernel, which we use later in the text. We introduce the matrix model measures in Section \ref{secMM}. In Section \ref{corrf} we define the correlation functions and show how to integrate over the angular part of the measure. In Section \ref{1p} we compute the one-point function, and in Section \ref{2-p} we perform the angular integration in the two-point functions in the general case. For measures of the form
\begin{equation}
\label{Potmeasure}
d\mu(P)= e^{-\tr V(P)} [dP]_{\rm Haar}\, ,
\end{equation}
where $V$ is a polynomial, we present the two-point function in a more explicit form in Section \ref{singletr} using the method of orthogonal polynomials. In Section \ref{intout} we derive another representation for the two-point function, which allows us to derive the general form of its large $k$ limit in Section \ref{Largek}. In Section \ref{ws} we focus on the Wishart model and derive the two-point function in an explicit form at finite $k$. We also compute its large $k$ limit explicitly, checking our previous general results.

\section{Background}
\label{background}

\subsection{Bergman metrics and Bergman kernel}

Here we collect some basic formulas from the setup of \cite{FKZ}, which we will use in this paper. Consider a compact \kahler manifold $M$ of complex dimension $n$. We fix a complex structure $J$ on $M$ and consider the set $\kcalomega$ of \kahler metrics in the cohomology class $[\omega_0]\in H^{1,1}(M,2\pi\mathbb Z)$, which can be parameterized by the space of \kahler potentials
\begin{equation}  \label{kcaldef} \kcal_{[\omega_0]} = \{\phi \in
C^{\infty}(M)/\R: \omega_{\phi} = \omega_0 + i \ddbar \phi > 0\}
\end{equation}
Under the integrality assumption on $[\omega_0]$, there exists a holomorphic line bundle $L\rightarrow M$ with $c_1(L)=[\omega_0]$. The space $\kcalomega$ is thus identified with the space $\hcal$ of Hermitian metrics $h=e^{-\phi}h_0$ on $L$ with the curvature $\omega_\phi=-\p\bp\log h\in[\omega_0]$. Consider then the $k$th power $L^k$ of the line bundle and the corresponding $N_k$-dimensional space $H^0(M,L^k)$ of holomorphic sections. We choose a basis of sections $\{s_i(z)\}=\bigl(s_1(z),...,s_{N_k}(z)\bigr)$, orthonormal with respect to the reference (background) metric $h_0^k$ on $L^k$,
\begin{equation}
\label{ortho}
\frac1V\int_M \bs_i(z)s_j(z)h_0^k\,\omega_0^n=\delta_{ij},
\end{equation}
where $\omega_0$ is the \kahler metric on $M$ corresponding to metric $h_0^k$ the line bundle
\begin{equation}
\omega_{0a\bb}=\frac1k\p_a\bp_{\bb}\log h_0^k.
\end{equation}
The basis of sections $\{s_i(z)\}$ is a vector in the linear space of dimension $N_k$ and it is defined up to a multiplication by a constant. Therefore it can be thought of as an embedding $z\rightarrow[s_1(z),...,s_{N_k}(z)]$ of $M$ into the projective space $\mathbb{CP}^{N_k-1}$ of sections (the Kodaira embedding). All other choices of bases $\{\tilde s_i(z)\}$ can be obtained by applying a linear transformation to the initial basis: $\tilde s_i(z)=A_{ij}s_j(z)$, for $A\in GL(N_k,\mathbb C)$. The set $\Bcal_k$ of Bergman metrics is then defined as the pullback of the Fubini-Study metric\footnote{With some abuse of notation we will make no distinction between the metric $g$ and the corresponding \kahler form $\omega$, connected as $\omega_{a\bb}=ig_{a\bb}$.} from $\mathbb{CP}^{N_k-1}$,
\begin{equation}
\label{bergm}
\omega_{\phi_P a\bb}(z)=\frac1k\p_a\bp_{\bb}\log \bs_i(z)P_{ij}s_j(z),
\end{equation}
where $P=A^{\dagger}A$ is a positive hermitian matrix in $GL(N_k,\mathbb C)/U(N_k)$.

Note that a rescaling of $P$ by a constant does not change the metric, so we can mod out by scalar matrices. Hence the  space  $\Bcal_k$ of Bergman metrics is equivalent to 
\begin{equation}
\label{symm}
\Bcal_k= SL(N_k,\mathbb C)/SU(N_k).
\end{equation}
The key fact, shown in \cite{Ti,Z,C}, is that for large $k$ the space of Bergman metrics approximates the space $\kcalomega$ of all \kahler metrics
\begin{equation}
\label{TYZth}
\kcalomega=\lim_{k\rightarrow\infty}\Bcal_k.
\end{equation}
The proof of (\ref{TYZth}) is based on the asymptotic expansion of the Bergman kernel on the diagonal \cite{Z,C,Lu,DKl}. For the orthonormal basis of sections as in Eq. (\ref{ortho}) the latter has the form 
\begin{equation}
\label{bk}
\rho_k(z)=|s(z)|^2h_0^k:=\sum_{i=1}^{N_k}\bs_i(z)s_i(z)h_0^k=k^n \bigl(1+\frac1{2k}R(\omega_0)+\ocal(1/k^2) \bigr).
\end{equation}
Here all the terms that appear on the right are various curvature invariants, depending only on the metric $\omega_0$.
The total number of sections $N_k$ is given by the integral of the Bergman kernel over the manifold. It grows at large $k$ as
\begin{equation}
N_k=k^n+\frac12c_1(M)k^{n-1}+\mathcal O(k^{n-2})
\end{equation}
see \cite{Lu} for the subleading terms.

\subsection{Off-diagonal Bergman kernel}

Another important function, which will appear later is the normalized off-diagonal Bergman kernel
\begin{equation}
\frac{|\langle\bs(z_1),s(z_2)\rangle|^2}{|s(z_1)|^2|s(z_2)|^2}:= \frac{1}{|s(z_1)|^2|s(z_2)|^2} \sum_{i,j=1}^{N_k}\bs_i(z_1)s_i(z_2)\bs_j(z_2)s_j(z_1).
\end{equation}
Note, that this is nothing but the squared norm of the inner product of the two vectors of sections $s_i(z_1)$ and $s_i(z_2)$. If $\theta_{12}$ is the angle between the two vectors, then the Bergman kernel is given by
\begin{equation}
\label{Szego}
\cos^2\theta_{12}(z_1,z_2)=\frac{|\langle\bs(z_1),s(z_2)\rangle|^2}{|s(z_1)|^2|s(z_2)|^2}.
\end{equation}
We will need the Taylor expansion of (\ref{Szego}) for small distances between the points $z_1,z_2$, which can be obtained following \cite{MM,LS,SZ}. First, we rewrite (\ref{Szego}) in terms of the Calabi's diastatic function, defined as follows. Any \kahler metric can be written in local coordinates as $\omega_{a\bb}=\p_a\bp_{\bb}\Phi(z,\bz)$, for some potential $\Phi(z,\bz)$. Consider $\Phi(z,\bz)$ formally as a function of two independent variables $z$ and $\bz$. Then the Calabi's diastatic function is defined as 
\begin{equation} 
D(z_1,\bz_2)=\Phi(z_1,\bz_1)+\Phi(z_2,\bz_2)-\Phi(z_1,\bz_2)-\Phi(z_2,\bz_1).
\end{equation}
This function is an invariant of the \kahler metric, independent of the choice of local coordinates \cite{Calabi}. Consider now the Bergman metric
\begin{equation}
\label{id}
\omega_{\phi_I}=\frac1k\p\bp\log|s(z)|^2,
\end{equation}
corresponding to the identity matrix $P=I$ \eref{bergm}. The normalized Bergman kernel (\ref{Szego}) can then be written as 
\begin{equation}
\label{kdependence}
\cos^2\theta_{12}(z_1,z_2)=e^{-kD_I(z_1,z_2)},
\end{equation}
where 
\begin{equation}
D_I(z_1,z_2)=\frac1k\bigl(\log|s(z_1)|^2+\log|s(z_2)|^2-\log|\langle\bs(z_1),s(z_2)\rangle|^2\bigr)
\end{equation}
is the diastatic function defined with respect to the metric (\ref{id}). This function equals zero only for coincident points $z_1,z_2$. In normal coordinates around a point $z_0=0$ we have the expansion
\begin{equation}
\Phi(z,\bz)=|z|^2+\frac14 R_{a\ba b\bb}(0)z^a\bz^{\ba}z^b\bz^{\bb}+\mathcal O(|z|^5)
\end{equation}
With some abuse of notation we assume that $z_1$ and $z_2$ are the coordinates of the points in the chosen local chart. Therefore we get the following Taylor expansion for small distances between $z_1$ and $z_2$
\begin{equation}
\label{diastasis}
\fl D_I(z_1,z_2)=|z_1-z_2|^2+\frac14R_{a\ba b\bb}(0)\left(z_1^a\bz_1^{\ba}z_1^b\bz_1^{\bb}+z_2^a\bz_2^{\ba}z_2^b\bz_2^{\bb}-z_1^a\bz_2^{\ba}z_1^b\bz_2^{\bb}-z_2^a\bz_1^{\ba}z_2^b\bz_1^{\bb}\right)+\ldots
\end{equation}
where all the quantities here are defined with respect to the metric (\ref{id}).
The expansion of the off-diagonal Bergman kernel follows immediately from the above formula.

\section{Eigenvalue measures on Bergman metrics}

\subsection{General form of the measure}
\label{secMM}

Many aspects of the infinite-dimensional geometry of $\kcalomega$ \eref{kcaldef} can be approximated by the symmetric space geometry of $\Bcal_k$ \eref{symm}. This is the main idea behind the Yau-Tian-Donaldson program in \kahler geometry \cite{Ti,Don,PS2}. In Ref. \cite{FKZ} we proposed to define probability measures  $d\mu_{\phi}$ on random \kahler metrics as large $k$ limit of probability measures on the finite-dimensional symmetric spaces $\Bcal_k$,
\begin{equation} 
\int_{\kcalomega}F(\omega_\phi)d\mu_{\phi}:=\lim_{k\rightarrow\infty}\int_{\Bcal_k}F_k(\omega_{\phi_P})d\mu_{\Bcal_k}(P).
\end{equation}
Here $F(\omega_\phi)$ schematically denotes operators, whose correlation functions we want to compute. Some examples of physically interesting measures $d\mu_{\phi}$ are described in \cite{FKZ}.

In particular, the natural metric on $\kcalomega$, the Mabuchi-Semmes-Donaldson metric, arises in the large $k$ limit from the invariant metric on $\Bcal_k$ \cite{CS}, see \cite{FKZ} for the details. The volume element of the invariant metric is the Haar measure
\begin{eqnarray} 
\label{HAAR} &&[dP]_{\rm Haar} = \frac{1}{(\det
P)^{N_k}}\,[dP]\\
&&\nonumber
[dP]= \frac{1}{(\det
P)^{N_k}}\; d P_{11} \cdots dP_{N_kN_k} \prod_{1\leq  i< j\leq N_k} d\, \Re
P_{ij}\, d\, \Im P_{ij}\,,
\end{eqnarray}
which we use as a basic building block for the random measure on the space of Bergman metrics. Recall for future reference the standard eigenvalue decomposition of the measure 
\begin{equation}
\label{decomp}
[dP]=\Delta^2(\lambda)\,[d\lambda][dU],\quad\Delta(\lambda)=\prod_{i<j}(\lambda_i-\lambda_j),\quad[d\lambda]=\prod_{i=1}^{N_k}d\lambda_i,
\end{equation}
where $P=U^\dagger\Lambda U$, $[dU]$ is the Haar measure on $U(N_k)$ and the eigenvalues are positive definite $\lambda_i>0$.

Simplest random matrix measures on positive hermitian matrices are the eigenvalue type measures. These are invariant under the $U(N_k)$ conjugation $P\rightarrow U^\dagger PU$. Therefore they have the general form of the product of the Haar measure (\ref{HAAR}) and an eigenvalue-dependent weight
\begin{equation}
\label{eig}
d\mu(P)= \mathcal F(\lambda) [dP]_{\rm Haar},
\end{equation}
where we use the standard decomposition of the hermitian matrix $P=U^\dagger \Lambda U$ into eigenvalue and angular part. We will also assume that $\mathcal F(\lambda) $ is a completely symmetric function of the eigenvalues.
The measure on $\Bcal_k$ can be obtained by gauge-fixing the scale invariance of Bergman metrics under $P\rightarrow c\cdot P$. This can be done by constraining the measure (\ref{eig}) to matrices $P$ with unit determinant. We thus in practice will consider the gauge-fixed measures $d\mu_{\Bcal_k}(P)$, defined as
\begin{equation}
\label{constm}
d\mu_{\Bcal_k}(P):=\mathcal F_{\Bcal_k}(\lambda)[dP]_{\rm Haar},
\end{equation}
where the new eigenvalue weight $\mathcal F_{\Bcal_k}$ on $\Bcal_k$ equals $\mathcal F$ multiplied by the delta-function constraint 
\begin{equation}
\mathcal F_{\Bcal_k}(\lambda) :=\mathcal F(\lambda)\delta\left(\sum_{i=1}^{N_k}\log\lambda_i\right).
\end{equation}
Note, that resulting measures are still strictly speaking of the form (\ref{eig}). The normalization is always chosen according to
\begin{equation}
\label{norm}
\int_{\Bcal_k}d\mu_{\Bcal_k}(P)=1.
\end{equation}
Another useful parametrization of the Bergman metrics is in terms of $GL(N_k,\mathbb C)$ matrix $A$, as 
\begin{equation}
\label{GLA}
P=A^\dagger A,
\end{equation}
with residual symmetry $A\rightarrow VA$, where $V\in U(N_k)$. The measure $[dP]$ can be represented as $[dP]=[dA]/[dV]$, where $[dA]$ is the Lebesgue measure on complex matrices 
\begin{equation}
\label{gln}
[dA]=\prod_{1\leq i,j \leq N_k} d\,{\rm Re}\,A_{ij}\,d\,{\rm Im}\,A_{ij}, 
\end{equation}
and $[dV]$ is the Haar measure on $ U(N_k)$.

\subsection{Correlation functions}
\label{corrf}

The basic set of correlation functions consists of the products of Bergman metrics at different points
\begin{eqnarray}
\label{corr}\nonumber
&&\E_k \,\omega_{\phi_P a_1\bb_1}(z_1)\ldots\omega_{\phi_P a_m\bb_m}(z_m):=\\\nonumber&&=\int_{\Bcal_k} \omega_{\phi_Pa_1\bb_1}(z_1) \ldots\omega_{\phi_P a_m\bb_m}(z_m)d\mu_{\Bcal_k}(P)\\
&&=\p_{a_1}\bp_{\bb_1} \ldots\p_{a_m}\bp_{\bb_m}\int_{\Bcal_k} \phi_P(z_1)...\phi_P(z_m) d\mu_{\Bcal_k}(P)
\end{eqnarray}
where we introduced the function
\begin{equation}
\label{berpot}
\phi_P(z)=\frac1k\log \bs_i(z) P_{ij}s_j(z),
\end{equation}
called the Bergman potential. It would also be interesting to consider more complicated geometric correlation functions, e.g.\ involving the Ricci tensor $R_{a\bb}=-\p_a\bp_{\bb}\log \omega_\phi^n$. Following \cite{FKZ0,FKZ}, the main idea is that in the large $N_k$ (equivalently, large $k$) limit matrix model correlation functions  on $\Bcal_k$ will tend to exact correlation functions for corresponding limiting random measures on the full space of \kahler metrics $\kcalomega$ 
\begin{equation}
\label{definition}
\big\langle\omega_{\phi a_1\bb_1}(z_1)\ldots\omega_{\phi a_m\bb_m}(z_m)\big\rangle:=\lim_{k\rightarrow\infty}\E_k \,\omega_{\phi_P a_1\bb_1}(z_1)\ldots\omega_{\phi_P a_m\bb_m}(z_m).
\end{equation}

For the eigenvalue models correlation functions depend explicitly on a choice of basis of sections, and therefore on the background metric $\omega_0$ (\ref{ortho}), corresponding to that choice. Note, that dependence of the Bergman potential \eref{berpot} on the vector $s_i$ breaks the $U(N_k)$ invariance of the integral, making the integration over the angular variables nontrivial. Luckily, for the models of the type (\ref{eig}) the integration over the angular part can be performed explicitly, using the Harish-Chandra-Itzykson-Zuber integral \cite{HC,IZ}. 
Indeed, we can use the following integral representation of logarithm
\begin{equation}
\log\alpha=\frac1t+\gamma-\int_0^{\infty} x^{t-1}e^{- \alpha x}dx+\mathcal O(t),
\end{equation}
for small $t$, and $\gamma$ is the Euler constant. Since the first two terms here are independent of  $z_i$, we can express the correlator (\ref{corr}) as follows
\begin{eqnarray}
\label{mi}
\fl \E_k \,\omega_{\phi_Pa_1\bb_1}(z_1)\ldots\omega_{\phi_P a_m\bb_m}(z_m)
=\\
\fl \nonumber=\frac{(-1)^m}{k^m}\p_{a_1}\bp_{\bb_1} \ldots\p_{a_m}\bp_{\bb_m}\lim_{\{t_p\}\rightarrow 0}\,\prod_{p=1}^m\int_0^\infty dx_p\,x_p^{t_p-1} \,\int_{\Bcal_k}e^{-\sum_{q=1}^{m}x_q\bs(z_q) Ps(z_q)}d\mu_{\Bcal_k}(P).
\end{eqnarray}
Assuming the ansatz (\ref{eig}) and using the eigenvalue decomposition (\ref{decomp}), the matrix integral in Eq. (\ref{mi}) reads
\begin{eqnarray}
\label{corr1}
&&\nonumber\int_{\Bcal_k}e^{-\sum_{q=1}^{m}x_q\bs(z_q) Ps(z_q)}d\mu_{\Bcal_k}(P)=
\\&&=\int_{\mathbb R_+^{N_k}}\int_{U(N_k)}e^{-\tr\, U\Lambda U^\dagger\bigl(\sum_{q=1}^{m}x_qs(z_q)\cdot\bs(z_q)\bigr)}\mathcal F_{\Bcal_k}(\lambda)\Delta^2(\lambda)\,[d\lambda][dU],
\end{eqnarray}
and the unitary integral here is of the Harish-Chandra-Itzykson-Zuber type
\begin{equation}
\label{HCIZ}
\int_{U(N)}e^{\tr \,UAU^\dagger B} \frac{[dU]}{{\rm Vol}\,U(N)}=\left(\prod_{p=1}^{N-1}p!\right)
\frac{\det\left(e^{a_i b_j}\right)_{1\leq i,j\leq N}}{\Delta(a)\Delta(b)}.
\end{equation}
Here $a_i$ and $b_j$ are the eigenvalues of some hermitian matrices $A$ and $B$. Therefore, the angular integral in Eq. (\ref{corr1}) depends on $\lambda_i$ and on the eigenvalues of the following matrix
\begin{equation}
\label{Phi}
{\Phi^{(m)}}_{ij}=\sum_{q=1}^{m}x_qs_i(z_q)\bs_j(z_q).
\end{equation}
This matrix is degenerate and has in general $m$ non-zero eigenvalues.

\subsection{One-point function}
\label{1p}
The formula (\ref{HCIZ}) simplifies further for small $m$, i.e.\ for small number of eigenvalues of $\Phi^{(m)}$ \eref{Phi}. In particular, the expectation value of a single metric can be found exactly for all eigenvalue type measures \eref{eig}. The matrix $\Phi^{(1)}$ has a single non-zero eigenvalue, which is equal to $x|s(z)|^2$, and the angular integral in (\ref{corr1}) gives
\begin{equation}
\fl \int_{U(N_k)}e^{-\tr\, U\Lambda U^\dagger\Phi^{(1)}} \frac{[dU]}{{\rm Vol}\,U(N_k)}=
\frac{(N_k-1)!}{\Delta(\lambda)\bigl(x|s(z)|^2\bigr)^{N_k-1}}\sum_{i=1}^{N_k}(-1)^{i-1}\Delta_i(\lambda)e^{-x|s(z)|^2\lambda_i},
\end{equation}
where $\Delta_i(\lambda)=\prod_{j<l,\,j,l\neq i}(\lambda_j-\lambda_l)$ is the Vandermonde determinant with omitted $i$'th eigenvalue.
We get for the expectation value of a single Bergman metric
\begin{eqnarray}
\nonumber
\fl \E_k \, \omega_{\phi_P}(z)=-\frac1k\p\bp \lim_{t\rightarrow 0}\,\int_0^\infty dx \,x^{t-1}\,\int_{\Bcal_k}e^{-x\bs(z) Ps(z)}d\mu_{\Bcal_k}(P)=-\frac1k\p\bp \lim_{t\rightarrow 0}|s(z)|^{-2t} \cdot\\
\fl \nonumber\cdot(N_k-1)!\int_0^\infty dx \,x^{t-N_k} \int_{\mathbb R_+^{N_k}}
\sum_{i=1}^{N_k}(-1)^{i-1}\Delta_i(\lambda)e^{-x\lambda_i}\mathcal F_{\Bcal_k}(\lambda)\Delta(\lambda)\,[d\lambda]\cdot{\rm Vol}\,U(N_k)\\
 \label{corr2}
=-\frac1k\p\bp \lim_{t\rightarrow 0}|s(z)|^{-2t}(N_k-1)!(-1)^{N_k-1}c_{N_k}\int_0^\infty\, x^{t-N_k}e^{-x}dx\\
 \nonumber = c_{N_k}\frac1k\p\bp \log|s(z)|^2=c_{N_k}\omega_{\phi_I}(z),
\end{eqnarray}
where the Bergman metric $\omega_{\phi_I}$ corresponds to the identity matrix $P=I$ (\ref{id}). We also sometimes omit the space indices, we it cannot lead to confusion.
The  constant $c_{N_k}$ equals one due to the normalization condition (\ref{norm}) for the measure 
\begin{equation}
\fl c_{N_k}=\int_{\mathbb R_+^{N_k}}
\sum_{i=1}^{N_k}(-1)^{N_k+i}\Delta_i(\lambda) \lambda_i^{N_k-1}\mathcal F_{\Bcal_k}(\lambda)\Delta(\lambda)\,[d\lambda] \cdot{\rm Vol}\,U(N_k)=\int_{\Bcal_k}d\mu_{\Bcal_k}=1,
\end{equation}
since the sum over eigenvalues here is just the expansion of the Vandermonde determinant along a column. Finally, using the asymptotic expansion (\ref{bk}), we get
\begin{equation}
\omega_{\phi_I}(z)=\omega_0(z)+\frac1k\p\bp\log\rho_k(z)=\omega_0(z)+\mathcal O(1/k^2),
\end{equation}
where $\omega_0$ is the background metric (\ref{ortho}). Therefore, in the large $k$ limit the one point function of the Bergman metric is equal to the background metric 
\begin{equation}
\label{corr3}
\big\langle\omega_\phi(z)\big\rangle= \lim_{k\rightarrow\infty}\E_k \, \omega_{\phi_P}(z)=\omega_0(z).
\end{equation}
This result is universal for models of the type (\ref{eig}). However, the case of the two-point function is model dependent. In the remainder of this section we analyze the two-point function for generic eigenvalue models, and in the Section \ref{ws} we derive it explicitly in the case of Wishart measure.

\subsection{Two-point function}
\label{2-p}

For the two-point function 
the matrix $\Phi^{(2)}$ (\ref{Phi}) has two non-zero eigenvalues
\begin{eqnarray}
\label{phieig} \nonumber
\phi_{1,2}&=&\frac12\bigl(x_1|s(z_1)|^2+x_2|s(z_2)|^2\bigr)\pm\\
&\pm&\frac12\sqrt{(x_1|s(z_1)|^2-x_2|s(z_2)|^2)^2+4\bigl|\langle\bs(z_1),s(z_2)\rangle\bigr|^2x_1x_2}.
\end{eqnarray}
Performing the angular integral in (\ref{corr1}) and rescaling $x_{1,2}\rightarrow x_{1,2} |s(z_{1,2})|^{-2}$, we get for the two-point function
\begin{eqnarray}
\label{general}\nonumber
\fl \E_k \, \omega_{\phi_P}(z_1)\omega_{\phi_P}(z_2)=\\\nonumber
\fl =\frac1{k^2}\p\bp|_{z_1}\p\bp|_{z_2}
\lim_{t_1,t_2\rightarrow 0}|s(z_1)|^{-2t_1}|s(z_2)|^{-2t_2} \int_0^\infty dx_1 \,x_1^{t_1-1} \int_0^\infty dx_2 \,x_2^{t_2-1}\\
\fl \cdot\frac{N_k!(N_k-1)!}{(\psi_1\psi_2)^{N_k-2}(\psi_1-\psi_2)}\int_{\mathbb R_+^{N_k}} \Delta(\lambda)\Delta_{12}(\lambda)e^{-\psi_1\lambda_1- \psi_2\lambda_2} \mathcal F_{\Bcal_k}(\lambda)[d\lambda] \cdot{\rm Vol}\,U(N_k),
\end{eqnarray}
where $\Delta_{12}$ is the Vandermonde determinant for eigenvalues $\lambda_{3},\lambda_{4},\ldots\lambda_{N_k}$, and we replaced (\ref{phieig}) by the rescaled eigenvalues 
\begin{equation}
\psi_{1,2}=\frac12\bigl(x_1+x_2\pm\sqrt{(x_1-x_2)^2+4\cos^2\theta_{12}x_1x_2}\bigr).
\end{equation}
Now we can see that all nontrivial coordinate dependence of the two-point function is contained in the norms $|s(z_i)|$ and in the off-diagonal Bergman kernel kernel $\cos^2\theta_{12}(z_1,z_2)$ (\ref{Szego}). From the expression (\ref{general}) we can read off the general structure of the two-point function. The small $t$ expansion of the  $x_1$, $x_2$-integrals in Eq. (\ref{general}) starts in general with singular terms 
\begin{equation}
\nonumber
\frac1{t_1t_2}+c\frac1{t_1}+c\frac1{t_2}+f_k(\cos^2\theta_{12}(z_1,z_2))+\mathcal O(t_1,t_2),
\end{equation}
and the coefficients in front of the singular terms are independent of the coordinates on the manifold.
Therefore the two-point function has schematically the following form
\begin{equation}
\label{2pf}
\fl \E_k \, \omega_{\phi_P}(z_1)\omega_{\phi_P}(z_2)= \omega_{\phi_I}(z_1)\omega_{\phi_I}(z_2)+
\frac1{k^2}\p\bp|_{z_1}\p\bp|_{z_2}f_k(\cos^2\theta_{12}(z_1,z_2)),
\end{equation}
and the scaling behavior of the function $f_k$ is to be determined.

\subsection{Single-trace potentials}
\label{singletr}

The two-point function (\ref{general}) is simplified further when the eigenvalue weight in Eq. (\ref{eig}) has the form of the exponent of a single-trace potential
\begin{equation}
\mathcal F(\lambda)=e^{-\tr V(\lambda)}
\end{equation}
The expression (\ref{general}) can be simplified using the method of orthogonal polynomials \cite{Mehta,DGZ}. Suppose $\{p_n(\lambda)\}$, $n=0,\ldots,N_k-1$ is the basis of orthogonal polynomials satisfying
\begin{equation}
\label{poly}
\int_0^\infty\, p_n(\lambda)p_m(\l)e^{-V(\l)} \l^{a}d\lambda =h_n(a)\delta_{nm}
\end{equation}
and normalized according to $p_n(\l)=\l^n+\ldots$ The parameter $a$ here is assumed to be pure imaginary and is used for the gauge-fixing. Indeed, we can connect the measures \eref{eig} and \eref{constm} as 
\begin{equation}
d\mu_{\Bcal_k}(P)=\int_{-i\infty}^{i\infty}da\, (\det P)^a d\mu(P).
\end{equation}
Then, using the relation
\begin{equation}
\label{detorth}
\Delta(\l)=\det P_{n-1}(\l_m),
\end{equation}
the normalization condition (\ref{norm}) reads
\begin{equation}
\int_{\Bcal_k}d\mu_{\Bcal_k}(P)=N_k!\int_{-i\infty}^{i\infty}da\prod_{n=0}^{N_k-1}h_n(a)\cdot{\rm Vol}\,U(N_k)=1.
\end{equation}
Let us now return to Eq.\ (\ref{general}) and rewrite the Vandermonde determinants in the integrand using Eq.\ (\ref{detorth}). 

The orthogonality condition Eq.\ \eref{poly} then allows us to perform all but two eigenvalue integrals, and we get
\begin{eqnarray} 
\label{2pforth}
\nonumber
&&\E_k \, \omega_{\phi_P}(z_1)\omega_{\phi_P}(z_2)=\\\nonumber
&&\frac1{k^2}\p\bp|_{z_1}\p\bp|_{z_2}
\lim_{t_1,t_2\rightarrow 0}|s(z_1)|^{-2t_1}|s(z_2)|^{-2t_2} \int_0^\infty dx_1 \,x_1^{t_1-1} \int_0^\infty dx_2 \,x_2^{t_2-1} \\\nonumber&& \cdot\frac{N_k!(N_k-1)!(N_k-2)!}{(\psi_1\psi_2)^{N_k-2}(\psi_1-\psi_2)}\int_{-i\infty}^{+i\infty}da\prod_{n=0}^{N_k-3}h_n(a) \cdot{\rm Vol}\,U(N_k) \\\nonumber&& \cdot\int_0^\infty d\lambda_1\,\l_1^a\int_0^\infty d\lambda_2\,\l_2^a\, e^{-V(\lambda_1)-V(\lambda_2)-\psi_1\lambda_1- \psi_2\lambda_2}\\&&\cdot\bigl(p_{N_k-2}(\lambda_1)p_{N_k-1}(\lambda_2)-p_{N_k-1}(\lambda_1)p_{N_k-2}(\lambda_2)\bigr).
\end{eqnarray}
We will use this representation later in order to derive the two-point function in the case of the Wishart potential.

\section{Integrating out matrix subspaces}
\subsection{One and two-point functions}
\label{intout}

Here we use the fact, that the matrix $\Phi^{(m)}$ in Eq. (\ref{Phi}) has only $m$ eigenvalues, in order to obtain another representation of the two-point function, useful for the large $k$ analysis. The idea is  to integrate out subspaces of $\Bcal_k$, corresponding to the directions where $\Phi^{(m)}$ has zero eigenvalues. The resulting formula will allow us to compute the general form of the large $k$ limit for measures of the type (\ref{eig}). 

In order to illustrate the idea, let us start first with the one-point function, and use explicitly the decomposition of the matrix $P$ into eigenvalues and angular variables
\begin{eqnarray}  
\E_k \, \omega_{\phi_P}(z)=\frac1k\p\bp  \int_{\Bcal_k}\log \bs_i(z)U^\dagger_{ij}\lambda_jU_{jl}s_l(z)\,d\mu_{\Bcal_k}(P).
\end{eqnarray}
Consider now the unit vector
$s_i(z)/|s(z)|\in \C^{N_k}$, and let $U_z \in U(N_k)$ be any unitary matrix such
that $(U^z {\bf e}_1)_i = s_i(z)/|s(z)|\in \C^{N_k}$, where ${\bf e}_1$ is the unit vector pointing in the direction $i=1$.
We get
\begin{eqnarray}  
&&\nonumber\E_k \, \omega_{\phi_P}(z)=\omega_{\phi_I}(z)+\frac1k\p\bp  \int_{\Bcal_k}\log
(UU^z{\bf e}_1)^\dagger_j\lambda_j(UU^z{\bf e}_1)_j\,d\mu_{\Bcal_k}(P)\\
&&=\nonumber\omega_{\phi_I}(z)+\frac1k\p\bp  \int_{\Bcal_k}\log
(U{\bf e}_1)^\dagger_j\lambda_j(U{\bf e}_1)_j\,d\mu_{\Bcal_k}(P)=\omega_{\phi_I}(z),
\end{eqnarray}
where in the second line we use the $U(N_k)$-invariance of the measure, and the resulting integral is independent of $z$. This is a short way to reproduce the result (\ref{corr2}).

Now we would like to calculate the two-point function
\begin{equation}
\label{2-pf}
\fl \E_k\, \phi_P(z_1)\phi_P(z_2)=
\frac1{k^2}\int_{\Bcal_k}\log\bs_i(z_1)U^\dagger_{ij}\lambda_jU_{jl}s_l(z_1)
\log\bs_i(z_2)U^\dagger_{ij}\lambda_jU_{jl}s_l(z_2)\,d\mu_{\Bcal_k}(P)
\end{equation}
using the same idea.
We can find a matrix $U^{z_1,z_2}$ such that
\begin{eqnarray}
&&\nonumber(U^{z_1,z_2} {\bf e}_1)_i = \frac{s_i(z_1)}{|s(z_1)|},\\&&\bigl( U^{z_1,z_2} ( {\bf e}_1\cos
\varphi(z_1,z_2) +  {\bf e}_2\sin \varphi(z_1,z_2))\bigr)_i = \frac{s_i(z_2)}{|s(z_2)|},
\end{eqnarray}
and taking the inner product of these two vectors we  immediately get
\begin{equation}
\cos^2 \varphi(z_1,z_2)=\frac{|\langle\bs(z_1),s(z_2)\rangle|^2}{|s(z_1)|^2|s(z_2)|^2} =\cos^2\theta_{12}(z_1,z_2)
\end{equation}
the off-diagonal Bergman kernel (\ref{Szego}). Note that there may exist many such matrices $U^{z_1,z_2}$. Using the $U(N_k)$ invariance of the measure, the integral (\ref{2-pf}) simplifies
\begin{eqnarray}
\label{block} \nonumber
\fl \E_k\, \phi_P(z_1)\phi_P(z_2)=\phi_I(z_1)\phi_I(z_2)+
\frac1{k^2}\int_{\Bcal_k}\log\sum_{i=1}^{N_k}\lambda_i|(UU^{z_1,z_2}{\bf e}_1)_i|^2
\\\nonumber
\fl \cdot\log \sum_{i=1}^{N_k}\lambda_i \bigl|(UU^{z_1,z_2} ( {\bf e}_1\cos
\theta_{12} +  {\bf e}_2\sin \theta_{12}))_i\bigr|^2 \,d\mu_{\Bcal_k}(P)= \phi_I(z_1)\phi_I(z_2) +\\
\fl  \nonumber +\frac1{k^2}\int_{\Bcal_k}\log \sum_{i=1}^{N_k}\lambda_i|(U{\bf e}_1)_i|^2\log \sum_{i=1}^{N_k}\lambda_i \bigl|(U( {\bf e}_1\cos
\theta_{12} + {\bf e}_2 \sin \theta_{12}))_i\bigr|^2 \,d\mu_{\Bcal_k}(P).
\end{eqnarray}
The last line can be expressed in terms of the matrix elements of $P$
\begin{eqnarray}
\fl \E_k\, \phi_P(z_1)\phi_P(z_2)= \phi_I(z_1)\phi_I(z_2) +\\
\label{2pf1}
\fl  \nonumber
+\frac1{k^2}\int_{\Bcal_k}\log P_{11}\log\bigl(P_{11}\cos^2
 \theta_{12}    +  (P_{12} + P_{21})\cos
 \theta_{12} \sin \theta_{12} +
 P_{22} \sin^2 \theta_{12}  \bigr)\,d\mu_{\Bcal_k}(P)
\end{eqnarray}
The integrand here depends only on the $2 \times 2$ block
\begin{eqnarray}
P^{(2)} : = \left(\begin{array}{cc}P_{11} & P_{12} \\P_{21} &
  P_{22} \end{array}\right) \in GL(2,\mathbb C)/U(2)=:\mathcal P_2.
\end{eqnarray}
One can check that $P^{(2)}$ is a positive Hermitian matrix since
for any $v \in \C^2$, $\langle P^{(2)} v, v \rangle = \langle P
(v, 0), (v, 0) \rangle$ where $(v, 0) \in \C^{N_k}$ is a vector
whose last $N_k-2$ entries are zero. Thus, we have the map
\begin{equation}
\label{projection}
\Pi_{2}: GL(N_k,\mathbb C)/U(N_k) \to GL(2,\mathbb C)/U(2)
\end{equation}  
which takes the upper left $2 \times 2$
block from $P$. Indeed, if $\Pi_2: \C^{d_k} \to \C^2$ is the map
$\Pi_2 (z_1, z_2, z') = (z_1, z_2, 0)$, then $P^{(2)} = \Pi_2 P
\Pi_2$ and it is clear that this is a positive semi-definite
Hermitian operator on $\C^{N_k}$ whose restriction to the range of
$\Pi_2$ is positive. We note also that $\pcal_2$ embeds in the
closure $\overline{\pcal_{N_k}}$ of positive semi-definite
Hermitian matrices if we put the zero matrix on the lower diagonal
$N_k-2$ block.

Since the integrand in Eq.\ (\ref{2pf1}) is manifestly a function of the three real variables $P_{11}$, $\Re P_{12}$, $P_{22}$, we have
\begin{eqnarray} 
\label{p-int}
\fl \int_{\Bcal_k}\log P_{11}
\log\bigl(P_{11}\cos^2
 \theta_{12}    +  (P_{12} + P_{21})\cos
 \theta_{12} \sin \theta_{12} +
 P_{22} \sin^2 \theta_{12}  \bigr)\,d\mu_{\Bcal_k}(P)\\ 
\fl  \nonumber =
\int_{\pcal_2} \log P_{11}\log\bigl(P_{11}\cos^2
 \theta_{12}    +  (P_{12} + P_{21})\cos
 \theta_{12} \sin \theta_{12} +
 P_{22} \sin^2 \theta_{12}  \bigr)\, d\mu_{\mathcal P_2,k}(P^{(2)}),
\end{eqnarray}
where $d\mu_{\mathcal P_2,k}(P^{(2)})$ is the pushforward measure on $\mathcal P_2$ under the projection (\ref{projection}),
\begin{equation}
\label{p2measure}
d\mu_{\mathcal P_2,k}(P^{(2)})=\left(\int \frac{d\mu_{\Bcal_k}(P)}{[dP^{(2)}]}\right)[dP^{(2)}],
\end{equation}
where  the integral on the right runs over all matrix elements of $P$ except those of $\mathcal P_2$ and $[dP^{(2)}]$ is the standard volume element on $\mathcal P_2$, given by
\begin{equation}
\label{haar2}
[dP^{(2)}]=dP_{11}\,dP_{22}\,d\, \Re
P_{12}\, d\, \Im P_{12}.
\end{equation}
The $U(2)$-invariance of this volume element immediately follows from (\ref{decomp}). We claim that $d\mu_{\mathcal P_2,k}(P^{(2)})$ is also $U(2)$ invariant under $P^{(2)}\rightarrow U^{(2)\dagger} P^{(2)} U^{(2)}$. Indeed, we may embed $U(2) \to U(N_k)$ as the upper $2 \times 2$ block, with the identity matrix  $I_{N_k - 2}$ on the lower diagonal. The original measure $d\mu_{\Bcal_k}(P)$ is $U(N_k)$ invariant and therefore is 
invariant under this subgroup. Then this fact and $U(2)$ invariance of (\ref{haar2}) imply $U(2)$ invariance of  $d\mu_{\mathcal P_2,k}(P^{(2)})$.

We may write the integral (\ref{p-int}) in the eigenvalue and polar coordinates as
follows. We write $P^{(2)} = U^{(2)\dagger} D^{(2)}(\lambda) U^{(2)}$ (non-uniquely), so that $U^{(2)} \in U(2), D^{(2)}(\lambda) = {\rm diag}(\lambda_1,
\lambda_2)$, ${\bf e}_1 = (1,0), {\bf e}_2 = (0, 1)$. Then the measure (\ref{p2measure}) has the  form
\begin{equation}
\label{p2form}
d\mu_{\mathcal P_2,k}(P^{(2)})=\mathcal F_{2,k}(\lambda) d\lambda_1d\lambda_2[dU^{(2)}],
\end{equation}
and $[dU^{(2)}]$ is the Haar measure on $U(2)$.
The function $\mathcal F_{2,k}(\lambda)$ here contains the information about the original measure on $\Bcal_k$. Now we can rewrite the integral  (\ref{p-int}) as  
\begin{eqnarray}
\nonumber
\fl \int_{\pcal_2} \log P_{11}\log\bigl(P_{11}\cos^2
 \theta_{12}    +  (P_{12} + P_{21})\cos
 \theta_{12} \sin \theta_{12} +
 P_{22} \sin^2 \theta_{12} \bigr)\, d\mu_{\mathcal P_2,k}(P^{(2)})=\\\nonumber
\fl \int_{\mathcal P_2}  \log \; | D^{(2)}(\lambda) U^{(2)} {\bf e}_1|^2
 \log | D^{(2)}(\lambda) U^{(2)}  ({\bf e}_1\cos
 \theta_{12}  + {\bf e}_2\sin \theta_{12})|^2\,[dU^{(2)}] \mathcal F_{2,k}(\lambda) d\lambda_1d\lambda_2.
  \end{eqnarray}
Consider first the inner integral here
\begin{eqnarray}
\label{i2int} 
\fl I_{2,k}(\lambda, \cos^2 \theta_{12})  =\\
\fl  \nonumber =
\int_{SU(2)} \log \; | D^{(2)}(\lambda) U^{(2)} {\bf e}_1|^2
 \log | D^{(2)}(\lambda) U^{(2)}  ({\bf e}_1\cos
 \theta_{12}  + {\bf e}_2\sin \theta_{12})|^2  [dU^{(2)}]. 
 \end{eqnarray}
 Note that scalar matrices in $U(2)$ commute with diagonal
matrices, and clearly cancel in the integrand, so the integral
over $U(2)$ is reduced to the one over $SU(2)$. We can parameterize the latter as 
 $$U = a_{\phi'} k_{\beta} a_{\phi}, $$
 where $$a_{\phi} = \left(\begin{array}{cc}e^{i \phi} & 0  \\0 & e^{- i \phi}\end{array}\right), \;\;\; k_{\beta} = \left(\begin{array}{cc}\cos \beta & \sin \beta\\- \sin \beta & \cos \beta\end{array}\right)$$
and the Haar measure on $SU(2)$ equals
\begin{equation}
[dU^{(2)}]=\frac1{8\pi^2}d\phi'\, d\phi\,\sin\beta\,d\beta,\quad 0\leq\phi,\phi'\leq2\pi\quad0\leq\beta\leq\pi.
\end{equation}
Since $a_{\phi'}$ commutes with $D^{(2)}(\lambda)$, the integrand in (\ref{i2int}) is independent of $\phi'$.  The first logarithm is also independent of $a_{\phi}$, and we can simplify the second logarithm in order to obtain the following integral, now over $S^2$
 \begin{eqnarray}
\fl \nonumber I_{2,k}(\lambda, \cos \theta_{12})=\\
\fl \nonumber  =
\frac1{4\pi}\int_{S^2} \log \; | D^{(2)}(\lambda)  k_{\beta} {\bf e}_1|^2
 \log | D^{(2)}(\lambda)  k_{\beta} a_{\phi}  ({\bf e}_1\cos
 \theta_{12}  + {\bf e}_2\sin \theta_{12})|^2\, d\phi\, \sin\beta\,d\beta\\
 \fl =
 \nonumber
\frac1{4\pi}\int_{S^2}\log \; | D^{(2)}(\lambda)  k_{\beta}  {\bf e}_1|^2
 \log | D^{(2)}(\lambda)  k_{\beta}   ({\bf e}_1e^{i\phi}\cos
 \theta_{12}  + {\bf e}_2e^{-i\phi}\sin \theta_{12})|^2\,d\phi\, \sin\beta\,d\beta. 
 \end{eqnarray}
Using the Jensen formula
$$\int_0^{\pi} \log (a + b \cos \phi) d\phi = \pi \log \frac{a
+ \sqrt{a^2 - b^2}}{2},\;\; a \geq |b| \geq 0 $$
we can perform the $\phi$-integration
\begin{eqnarray}
\nonumber
&&\int_0^{2\pi}
 \log | D^{(2)}(\lambda)  k_{\beta}   ({\bf e}_1e^{i\phi}\cos
 \theta_{12}  + {\bf e}_2e^{-i\phi}\sin \theta_{12})|^2   d \phi=\\\nonumber&&=\int_0^{2\pi} \log (A + B\cos 2 \phi) d\phi=2\pi\log \frac{A+\sqrt{A^2-B^2}}2,
 \end{eqnarray}
 where we introduced the following notations
\begin{eqnarray}
\fl \nonumber
A = |D^{(2)}(\lambda)  k_{\beta}{\bf e}_1|^2 \cos^2 \theta_{12}  + |D^{(2)}(\lambda)  k_{\beta} {\bf e}_2|^2 \sin^2 \theta_{12}= \\
\fl \nonumber=(\lambda_1^2 \cos^2 \beta + \lambda_2^2 \sin^2 \beta)\cos^2 \theta_{12}+(\lambda_1^2 \sin^2 \beta + \lambda_2^2 \cos^2 \beta)\sin^2 \theta_{12},\\
\fl \nonumber B =  2  \bigl( D^{(2)}(\lambda)  k_{\beta} {\bf e}_1, D^{(2)}(\lambda)  k_{\beta} {\bf e}_2 \bigr) \cos \theta_{12} \sin \theta_{12} =2(\lambda_1^2 - \lambda_2^2)\cos \theta_{12} \sin \theta_{12} \cos \beta\sin \beta .
\end{eqnarray}

This removes one integral and leaves us with
 \begin{equation} 
 \label{I2}
\fl I_{2,k}(\lambda, \cos^2 \theta_{12})  =
\frac12\int_0^\pi \log (\lambda_1^2 \cos^2 \beta + \lambda_2^2 \sin^2 \beta)
 \log \frac{A
+ \sqrt{A^2 - B^2}}{2}\, \sin\beta\,d\beta. \end{equation}
Our final answer for the two-point function (\ref{2-pf}) is given by
\begin{equation}
\label{final2p}
\fl \E_k\, \phi_P(z_1)\phi_P(z_2)=\phi_I(z_1)\phi_I(z_2)+
\frac1{k^2}\int_{\mathbb R_+^2} I_{2,k}(\lambda, \cos^2 \theta_{12}) \mathcal F_{2,k}(\lambda) d\lambda_1d\lambda_2,
\end{equation}
where the function $\mathcal F_{2,k}$ is defined in (\ref{p2form}). Now we are ready to perform the analysis of the large $k$ of the two-point function.

\subsection{Large $k$ limit in general case}
\label{Largek}

Representation (\ref{final2p}) allows us to study the large $k$ limit of the two-point correlation function for measures of the type (\ref{eig}). Note that the main coordinate dependence of the two-point function is encoded in the normalized off-diagonal Bergman kernel, which depends exponentially on $k$ (\ref{kdependence}). Therefore we distinguish the following two cases when considering the large $k$ limit:

(1) $kD_I(z_1,z_2) \rightarrow \infty$, or equivalently $\cos^2\theta_{12}\rightarrow0$ as $k\rightarrow\infty$, 

(2) $kD_I(z_1,z_2)\rightarrow 0$, or equivalently $\cos^2\theta_{12}\rightarrow1$ as $k\rightarrow\infty$. 

In the case (1) the distance between the points $z_1$ and $z_2$ is finite and does not scale with $k$, so we can use $\zeta=\cos^2\theta_{12}$ as a small expansion parameter. In this case we obtain a limiting expression outside the diagonal $z_1=z_2$.
In the case (2) the distance between $z_1$ and $z_2$ decreases effectively as $1/k^{1/2}$ or faster. Therefore, in this case $\epsilon=kD_I(z_1,z_2)$ becomes the small parameter. This limit will produce contact terms, supported on the diagonal $z_1=z_2$.

The key fact is that the integral (\ref{I2}) is well-behaved and admits a Taylor expansion both around $\zeta=\cos^2\theta_{12}=0$ and $\epsilon=kD_I(z_1,z_2)=0$. Our only assumption here is that this Taylor expansion is allowed to be used in the integral over eigenvalues in Eq. (\ref{final2p}). Therefore, in order to study the large $k$ limit in the case (1) we can expand the second logarithm in (\ref{I2}) around $\zeta=0$ 
\begin{equation}
\label{zeta0}
\fl I_{2,k}(\lambda, \cos^2 \theta_{12}) = I_{2,k}(\lambda, 0)+ \frac d{d\zeta}I_{2,k}(\lambda, \zeta)\big|_{\zeta=0}e^{-kD_I(z_1,z_2)}+\mathcal O(e^{-2kD_I(z_1,z_2)}),
\end{equation}
Where $I_{2,k}(\lambda, 0)$ and the first derivative at zero $I'_{2,k}(\lambda, 0)$ are given by certain convergent integrals, whose explicit form we do not need for this analysis. For finite $D_I(z_1,z_2)$ all the terms starting from the second on the right go to zero uniformly as $k\rightarrow\infty$. Therefore, we conclude that
\begin{equation}
\label{nondiag}
\fl \big\langle\omega_\phi(z_1)\omega_\phi(z_2)\big\rangle =\lim_{k\rightarrow\infty}\E_k\,\omega_P(z_1)\omega_P(z_2)=\omega_0(z_1)\omega_0(z_2),\quad D_I(z_1,z_2)>0.
\end{equation}

In the case (2) we use $kD_I(z_1,z_2)$ as the small parameter.  Since $D_I\sim |z_1-z_2|^2+\mathcal O(|z_1-z_2|^4)$, as follows from Eq. (\ref{diastasis}), and since the correlation function of two Bergman metrics contains four $z$-derivatives, it suffices to expand (\ref{I2}) up to $(kD_I)^2$ order
\begin{eqnarray}
\nonumber
\label{kexp}
\fl \E_k\,\omega_{\phi_Pa_1\bb_1}(z_1)\omega_{\phi_Pa_2\bb_2}(z_2)= \\
\fl  \nonumber=\omega_{\phi_Ia_1\bb_1}(z_1)\omega_{\phi_Ia_2\bb_2}(z_2)+\frac1{k^2}\lim_{z_1\rightarrow z_2}\p_{a_1}\bp_{\bb_1}\p_{a_2}\bp_{\bb_2} \int_{\mathbb R_+^2}\left( \frac d{d\epsilon}I_{2,k}(\lambda, e^{-\epsilon})\big|_{\epsilon=0}\,kD_I(z_1,z_2)+ \right.\\
\fl  \nonumber \left.+\frac12\frac{d^2}{d\epsilon^2}I_{2,k}(\lambda, e^{-\epsilon})\big|_{\epsilon=0}\,(kD_I(z_1,z_2))^2+\mathcal O((kD_I(z_1,z_2))^3) \right) \mathcal F_{2,k}(\lambda) d\lambda_1d\lambda_2= \\
\fl
=\omega_{\phi_Ia_1\bb_1}(z_1)\omega_{\phi_Ia_2\bb_2}(z_2)+c_{k} \bigl(\omega_{\phi_Ia_1\bb_1}\omega_{\phi_Ia_2\bb_2}+\omega_{\phi_Ia_2\bb_1}\omega_{\phi_Ia_1\bb_2}\bigr)\big|_{z_1}\delta_{z_1,z_2}
\end{eqnarray}
where $\delta_{z_1,z_2}$ equals one on the diagonal and zero outside. The constant $c_k$ is given by
\begin{equation}
\label{c-int}
c_{k}= \int_{\mathbb R_+^2} \frac{d^2}{d\epsilon^2}I_{2,k}(\lambda, e^{-\epsilon})\big|_{\epsilon=0}\, \mathcal F_{2,k}(\lambda) d\lambda_1d\lambda_2,
\end{equation}
and the final result depends on its asymptotic value
\begin{equation}
c=\lim_{k\rightarrow\infty}c_{k}.
\end{equation}
Combining the on-diagonal (\ref{kexp}) and off-diagonal (\ref{nondiag}) results and taking the large $k$ limit we get
\begin{eqnarray}
\label{res} \nonumber
\big\langle\omega_{\phi a_1\ba_1}(z_1)\omega_{\phi a_2\ba_2}(z_2)\big\rangle =\lim_{k\rightarrow\infty}\E_k\,\omega_{\phi_P a_1\ba_1}(z_1)\omega_{\phi_P a_2\ba_2}(z_2)=\\
=\omega_{0a_1\bb_1}(z_1)\omega_{0a_2\bb_2}(z_2)+c \bigl(\omega_{0a_1\bb_1}\omega_{0a_2\bb_2}+\omega_{0a_2\bb_1}\omega_{0a_1\bb_2}\bigr)\big|_{z_1}\delta_{z_1,z_2}.
\end{eqnarray}
which is the general form of the two-point correlation function for the eigenvalue type measures \eref{eig}.

Let us also note that in addition to the off and on-diagonal limits $(1)$ and $(2)$, one can also consider the scaling limits of the correlation function (\ref{final2p}). They correspond to the "near-diagonal" asymptotics of Ref. \cite{SZ} and reveal a fine structure behind the contact terms in (\ref{res}). The idea is to make the distance scale as $D^2_I(z_1,z_2)\leq b^2\log k/k$ for some constant $b$ as $k\rightarrow\infty$. In local coordinates around $z_0$ this amount to taking $z_1=z_0+u/\sqrt k$ and $z_2=z_0+v/\sqrt k$, where $|u|,|v|\leq b\sqrt{\log k}$. The Bergman kernel (\ref{kdependence}) then behaves as 
\begin{equation}
\cos^2\theta_{12}(z_0+u/\sqrt k,z_0+v/\sqrt k)\sim e^{-|u-v|^2},
\end{equation}
and the Taylor expansions of the previous cases no longer apply. Existence of such scaling limits is an intrinsic feature of the models of eigenvalue type.

\section{Wishart ensemble of random Bergman metrics}
\label{ws}

\subsection{Wishart distribution}

Now we would like to consider an explicit solvable example where correlation functions can be computed explicitly for finite $k$. 

Consider the Wishart distribution on positive hermitian $N_k\times N_k$ matrices, given by
\begin{equation}
\label{wish}
d\mu_a(P)=\frac1{Z_g} e^{-g\,\tr P} (\det P)^a[dP],
\end{equation}
and depending on the two parameters $a$ and $g$.
It can be equivalently written as in Eq. \eref{GLA} in terms of the $GL(N_k,\mathbb C)$ matrix $A$, taking $P=A^\dagger A$ with residual symmetry $A\rightarrow VA$
\begin{equation}
\label{WishartGL}
d\mu_a(A^\dagger A)=\frac1{Z_g}e^{-g\,\tr A^\dagger A}(\det A^\dagger A)^a\frac{[dA]}{[dV]},
\end{equation}
where the measure $[dA]$ is defined in Eq.\ (\ref{gln}). 
The constrained Wishart measure (\ref{constm}) on $\Bcal_k$ can be obtained from \eref{wish} by the Laplace transform
\begin{equation}
\label{wisbcal}
d\mu_{\Bcal_k}(P)=\int_{-i\infty}^{i\infty}d\mu_a(P) \,da=\frac1{Z_g} e^{-g\,\tr P} \delta(\log\det P)[dP]
\end{equation}
and we are left with a single parameter $g$.
The constant $Z_g$ is determined by the normalization condition (\ref{norm}), which reads
\begin{equation}
Z_g=\int_{\Bcal_k} e^{-g\tr P}\delta(\log\det P)[dP]
\end{equation}
More explicitly
\begin{eqnarray}
\label{Meijer}
\nonumber
&&Z_g=\int_{-i\infty}^{i\infty}da \,g^{-N_k(N_k+a)}\int_{\mathbb R_+^{N_k}}\,e^{-\sum_{i=1}^{N_k}\lambda_i}|\Delta(\lambda)|^2\prod_{i=1}^{N_k}\lambda_i^ad\lambda_i\cdot\vol\, U(N_k)=\\\nonumber &&=
\int_{-i\infty}^{i\infty}da \,g^{-N_k(N_k+a)}\prod_{j=1}^{N_k}
\Gamma (j+1)\Gamma(j+a) \cdot\vol\, U(N_k) =\\&&=g^{-N_k^2}\left[\prod_{i=1}^{N_k}
\Gamma (j+1)\right]\cdot G^{N_k,0}_{0,N_k}(g^{N_k}|1,2,...,N_k)\cdot \vol\, U(N_k),
\end{eqnarray}
where the Meijer $G$-function in the last line is defined by the integral in the second line. The integral in the second line is carried out using the Selberg formula, see e.g. Eq. (17.6.5) in Ref. \cite{Mehta}. We also define another normalization constant $Z_{g,a}$ as
\begin{equation}
Z_{g,a}=g^{-N_k(N_k+a)}\prod_{j=1}^{N_k}
\Gamma (j+1)\Gamma(j+a) \cdot\vol\, U(N_k) 
\end{equation}

\subsection{One-point function}

Correlation functions in the Wishart model can be computed using the $GL(N_k,\mathbb C)$ representation of the measure \eref{WishartGL}. Let us consider first the case of the one-point function.
We introduce the notation $\E^a_k$ for the expectation value with respect to the unconstrained measure (\ref{wish}), related to the expectation value $\E_k$ with respect to the constrained measure \eref{wisbcal} via the Laplace transform 
\begin{equation}
\label{exp}
\E_k \ldots=\int_{-i\infty}^{i\infty}da\,\E_k^a\ldots
\end{equation}
Then for the one-point function we get
\begin{equation}
\E_k^a\,\omega_{\phi_P}=-\frac1k\p\bp\lim_{t\rightarrow 0}\,\int_0^{\infty} dx\,x^{t-1}\E_k^a\,e^{-x\bs(z) Ps(z)},
\end{equation}
Since the expression in the exponent is linear in $P$ we can rewrite the expectation value on the right as
\begin{eqnarray}
\label{glrep}&& \nonumber
\E_k^a\,e^{-x\bs(z) Ps(z)}=\frac1{Z_g}\int e^{-g\,\tr P\bigl(I+\frac xgs\cdot\bs(z)\bigr)}\bigl(\det P\bigr)^a\,[dP]=\\&&=\frac1{Z_g}\int e^{-g\,\tr A^\dagger A\bigl(I+\frac xgs\cdot\bs(z)\bigr)}\bigl(\det A^\dagger A\bigr)^a\,\frac{[dA]}{[dV]},
\end{eqnarray}
where in the second line we use the $GL(N_k,\mathbb C)$ representation (\ref{WishartGL}).
The positive definite hermitian matrix in the exponent of Eq.\ (\ref{glrep}) can be written as
\begin{equation}
\label{b}
I_{ij}+ \frac xgs_i\cdot\bs_j=(BB^\dagger)_{ij}
\end{equation}
for some matrix $B\in GL(N_k,\mathbb C)$. 

Using the multiplication covariance property $[d (BA)]=(\det B^\dagger B)^{N_k} \,[dA]$, we immediately get
\begin{eqnarray}
\label{1pt}
\nonumber
\E_k^a\, \omega_{\phi_P}&=&-\frac1k\frac{Z_{g,a}}{Z_g}\,\p\bp\lim_{t\rightarrow0}\int_0^{\infty} dx\,x^{t-1}\bigl[\det(I+\frac xgs\cdot\bs)\bigr]^{-(N_k+a)}=\\\nonumber&=&
-\frac1k\frac{Z_{g,a}}{Z_g}\,\p\bp\lim_{t\rightarrow0}\int_0^{\infty} dx\,x^{t-1}
\bigl(1+\frac xg|s|^2\bigr)^{-(N_k+a)}dx=\\
&=&-\frac1k\frac{Z_{g,a}}{Z_g}\,\p\bp\lim_{t\rightarrow0}|s|^{-2t}g^t\,\mathrm B(N_k+a,t)=\frac1k\frac{Z_{g,a}}{Z_g}\,\p\bp\log |s|^2,
\end{eqnarray}
where $\mathrm B(N_k+a,t)$ is the beta-function, with the following behavior near $t=0$ 
$$
\mathrm B(t,N_k+a-t)=\frac{\Gamma(N_k+a-t)\Gamma(t)}{\Gamma(N_k+a)} =\frac1t+\mathcal O(1).
$$
The correlator (\ref{exp}) on $\Bcal_k$ agrees with the previous result (\ref{corr2}),
\begin{equation}
\E_k\, \omega_{\phi_P}(z)=\int_{-i\infty}^{i\infty} da\,\E_k^a\, \omega_{\phi_P}(z) =\frac1k\p\bp\log |s|^2=\omega_0(z)+\mathcal O(1/k^2).
\end{equation}

\subsection{Two-point function}

For the two-point function the calculation is analagous. We write
\begin{eqnarray}
\label{2pointf}
\fl \E_k^a\,\omega_{\phi_P}(z_1) \,\omega_{\phi_P}(z_2) =\\
\fl  \nonumber =
\frac1{k^2}\p\bp|_{z_1}\p\bp|_{z_2}\lim_{t_1,t_2\rightarrow0} \int_0^\infty dx_1\,x_1^{t_1-1}\int_0^{\infty}dx_2 \,x_2^{t_2-1}
\E_k^a\,e^{-\frac{x_1}g\bs(z_1) Ps(z_1)-\frac{x_2}g \bs(z_2)Ps(z_2)}.
\end{eqnarray}
The matrix $BB^\dagger$ (\ref{b}) in this case equals
\begin{equation}
BB^\dagger=I+\frac{x_1}gs(z_1)\cdot \bs(z_1) +\frac{x_2}g s(z_2)\cdot \bs(z_2)
\end{equation}
and the expectation value of the exponent reads
\begin{eqnarray}
\nonumber
\fl \E_k^a\,e^{-\frac{x_1}g\bs_1 Ps(z_1)-\frac{x_2}g \bs_2Ps(z_2)}=\\ \nonumber
\fl =\frac{Z_{g,a}}{Z_{g}}\bigl[\det(I+ \frac{x_1}g s(z_1)\cdot\bs(z_1)+ \frac{x_2}g s(z_2)\cdot\bs(z_2))\bigr]^{-(N_k+a)}=\\
\fl =\nonumber
 \frac{Z_{g,a}}{Z_{g}}\bigl(1+ \frac{x_1}g|s(z_1)|^2+ \frac{x_2}g|s(z_2)|^2+ \frac{x_1x_2}{g^2}(|s(z_1)|^2|s(z_2)|^2-|(\bs(z_1),s(z_2))|^2)\bigr)^{-(N_k+a)}.
\end{eqnarray}
Rescaling $x_{1,2}\rightarrow x_{1,2} |s(z_{1,2})|^{-2}/g$ we reduce the two-point function to
\begin{eqnarray}
\label{2pointf1}
\nonumber
&&\E_k^a\,\omega_{\phi_P}(z_1) \,\omega_{\phi_P}(z_2)=\\\nonumber
&& =\frac1{k^2} \frac{Z_{g,a}}{Z_{g}}\p\bp|_{z_1}\p\bp|_{z_2}\lim_{t_1,t_2\rightarrow0}
|s(z_1)|^{-2t_1}g^{t_1}|s(z_2)|^{-2t_2}g^{t_2}\\&&
\cdot\int_0^\infty dx_1\,x_1^{t_1-1}\int_0^{\infty}dx_2 \,x_2^{t_2-1}(1+x_1+x_2+ x_1x_2\sin^2\theta_{12})^{-(N_k+a)}.
\end{eqnarray}
We also derive this formula in the Appendix in a more direct way, using the method of orthogonal polynomials. 

Now the integrals over $x_1,x_2$ in \eref{2pointf1} can be performed
\begin{eqnarray}
\label{2point}
&&\E_k^a\,\omega_{\phi_P}(z_1) \,\omega_{\phi_P}(z_2)=\\\nonumber&&=\frac1{k^2} \frac{Z_{g,a}}{Z_{g}}\p\bp|_{z_1}\p\bp|_{z_2}\lim_{t_1,t_2\rightarrow0}
|s(z_1)|^{-2t_1}g^{t_1}|s(z_2)|^{-2t_2}g^{t_2} \cdot\\
&&\nonumber\mathrm B(t_1,N_k+a-t_1) \mathrm B(t_2,N_k+a-t_2)\,\phantom{}_2F_1(t_1,t_2,N_k+a;\cos^2\theta_{12}).
\end{eqnarray}
In order to compute the small $t$ limit in Eq. (\ref{2point}) we use the following representation of the hypergeometric function
\begin{eqnarray}
&&\nonumber
\mathrm B(t_2,N_k+a-t_2)\,\phantom{}_2F_1(t_1,t_2,N_k+a;\cos^2\theta_{12})=\\&&\nonumber=\int_0^1dx\,x^{t_2-1}(1-x)^{N_k+a-t_2-1}(1-x\cos^2\theta_{12})^{-t_1},
\end{eqnarray}
see e.g.\ Ref.\ \cite{BE}. Then for the expression in Eq. (\ref{2point}) we get
\begin{eqnarray}
\fl \nonumber \lim_{t_1,t_2\rightarrow0}
|s(z_1)|^{-2t_1}g^{t_1}|s(z_2)|^{-2t_2}g^{t_2}\,\mathrm B(t_1,N_k+a-t_1) \mathrm B(t_2,N_k+a-t_2)\\\fl  \nonumber\cdot\,\phantom{}_2F_1(t_1,t_2,N_k+a;\cos^2\theta_{12})=\\
\fl \nonumber=\lim_{t_1,t_2\rightarrow0}
|s(z_1)|^{-2t_1}g^{t_1}|s(z_2)|^{-2t_2}g^{t_2}\,\mathrm B(t_1,N_k+a-t_1)\\
\fl  \nonumber\cdot\int_0^1dx\,x^{t_2-1}(1-x)^{N_k+a-t_2-1}(1-x\cos^2\theta_{12})^{-t_1}=\\
\fl \nonumber=\lim_{t_1,t_2\rightarrow0}
|s(z_1)|^{-2t_1}g^{t_1}|s(z_2)|^{-2t_2}g^{t_2}B(t_1,N_k+a-t_1)\\
\fl \nonumber\cdot\left(\mathrm B(t_2,N_k+a-t_2)-t_1\int_0^1dx\,x^{t_2-1}(1-x)^{N_k+a-t_2-1}\ln(1-x\cos^2\theta_{12})\right)=\\
\fl \nonumber
=\lim_{t_1,t_2\rightarrow0}
|s(z_1)|^{-2t_1}g^{t_1}|s(z_2)|^{-2t_2}g^{t_2}\,\mathrm B(N_k+a,t_1) \mathrm B(N_k+a,t_2)
\\
\fl  \nonumber-\int_0^1\frac{dx}x(1-x)^{N_k+a-1}\ln(1-x\cos^2\theta_{12})
\end{eqnarray}
Plugging this back in Eq. (\ref{2point}) and taking the limit  we get
\begin{eqnarray}
\nonumber
\fl \E_k^a\,\omega_{\phi_P}(z_1) \,\omega_{\phi_P}(z_2)=\\
\fl \nonumber=\frac{Z_{g,a}}{Z_{g}} \omega_{\phi_I}(z_1)\omega_{\phi_I}(z_2)
-\frac{Z_{g,a}}{Z_{g}} \frac1{k^2}\p\bp|_{z_1} \p\bp|_{z_2}
\int_0^1\frac{dx}x(1-x)^{N_k+a-1}\ln(1-x\cos^2\theta_{12})
\end{eqnarray}
and the full expectation value (\ref{exp}) can be expressed using the $G$-function
\begin{eqnarray}
\label{answer}
\fl  \nonumber\E_k\,\omega_{\phi_P}(z_1) \,\omega_{\phi_P}(z_2)=\\
\fl  \nonumber
= \omega_{\phi_I}(z_1)\omega_{\phi_I}(z_2)-\frac1{k^2} \p\bp|_{z_1} \p\bp|_{z_2} \int_{-i\infty}^{i\infty}da\, \frac{Z_{g,a}}{Z_{g}}
\int_0^1\frac{dx}x(1-x)^{N_k+a-1}\ln(1-x\cos^2\theta_{12})=\\
\fl \nonumber
=\omega_{\phi_I}(z_1)\omega_{\phi_I}(z_2)-\\
\fl -\frac1{k^2} \p\bp|_{z_1} \p\bp|_{z_2}\int_0^1\frac{dx}x(1-x)^{N_k-1}\ln(1-x\cos^2\theta_{12})\,\frac{G^{N_k,0}_{0,N_k}(\frac{g^{N_k}}{1-x}|1,2,...,N_k)}{G^{N_k,0}_{0,N_k}(g^{N_k}|1,2,...,N_k)}
\end{eqnarray}
which is the final answer, valid at finite $k$. Now we proceed to the analysis of its large $k$ behavior.

\subsection{Large $k$ limit}
\label{w2p}

The integral (\ref{answer}) can be Taylor expanded around $\cos^2\theta_{12}=0$. Therefore, in the case (1) of Section \eref{Largek} of finite $D_I(z_1,z_2)>0$ the large $k$ behavior of the second term in (\ref{answer}) is given by
\begin{eqnarray}
\label{answer1} \nonumber
\fl -\frac1{k^2} \p\bp|_{z_1} \p\bp|_{z_2}\int_0^1\frac{dx}x(1-x)^{N_k-1}\ln(1-x\cos^2\theta_{12})\,\frac{G^{N_k,0}_{0,N_k}(\frac{g^{N_k}}{1-x}|1,2,...,N_k)}{G^{N_k,0}_{0,N_k}(g^{N_k}|1,2,...,N_k)}\approx\\\nonumber
\fl  \approx\frac1{k^2} \p\bp|_{z_1} \p\bp|_{z_2}e^{-kD_I(z_1,z_2)}+\mathcal O(e^{-2kD_I(z_1,z_2)}),
\end{eqnarray}
which is exactly the behavior predicted by Eq.\ (\ref{zeta0}).
In order to compute the contact terms, we expand (\ref{answer}) for small distances up to the second order in $kD_I(z_1,z_2)$
\begin{eqnarray} 
\label{answer2}\nonumber
\fl\E_k\,\omega_{\phi_Pa_1\bb_1}(z_1) \,\omega_{\phi_Pa_2\bb_2}(z_2) = \omega_{\phi_Ia_1\bb_1}(z_1)\omega_{\phi_Ia_2\bb_2}(z_2)-\\\fl  \nonumber
-\frac1{k^2} \p_{a_1}\bp_{\bb_1} \p_{a_2}\bp_{\bb_2}\int_1^\infty\frac{dy}{y^{N_k}}\bigl(kD-\frac12(kD)^2y\bigr) \frac{G^{N_k,0}_{0,N_k}(g^{N_k}y|1,2,...,N_k)}{G^{N_k,0}_{0,N_k}(g^{N_k}|1,2,...,N_k)}=\\
\fl
= \omega_{\phi_Ia_1\bb_1}(z_1)\omega_{\phi_Ia_2\bb_2}(z_2)
 +\\
\fl \nonumber
+\bigl(\omega_{\phi_Ia_1\bb_1}\omega_{\phi_Ia_2\bb_2}+\omega_{\phi_Ia_2\bb_1}\omega_{\phi_Ia_1\bb_2}\bigr)\big|_{z_1} \frac{G^{N_k,0}_{0,N_k}(g^{N_k}|1,2,...,N_k-2,N_k-2,N_k)}{G^{N_k,0}_{0,N_k}(g^{N_k}|1,2,...,N_k)} \delta_{z_1,z_2} ,
\end{eqnarray}
see Section 5.6.4 of Ref.\ \cite{Luke} for the relevant integrals of the Meijer $G$-function. Note, that the contact terms here have the same structure as in Eq.\ (\ref{res}).

The large $k$ limit value of the ratio of the $G$-functions in Eq.\ (\ref{answer2}) depends on the $k$ dependence of the parameter $g=g(k)$. If $g\rightarrow0$ as $k\rightarrow\infty$, then this ratio tends to one \cite{MN}, and we get 
\begin{eqnarray}
\fl\nonumber\big\langle\omega_{\phi a_1\ba_1}(z_1)\omega_{\phi a_2\ba_2}(z_2)\big\rangle_{g\rightarrow0}=\\=\nonumber\omega_{0 a_1\ba_1}(z_1)\omega_{0 a_2\ba_2}(z_2)+\bigl(\omega_{0a_1\ba_1}(z_1)\omega_{0a_1\ba_1}(z_1)+\omega_{0a_2\ba_1}(z_1)\omega_{0a_1\ba_2}(z_1)\bigr)\delta_{z_1,z_2},
\end{eqnarray}

The ratio of the $G$-functions in (\ref{answer2}) is also known if the parameter $g$ scales as $g\sim N_k^{3+ \varepsilon},\, \varepsilon >0$, see e.g.\ Ref.\ \cite{Fields}. In this case there is an asymptotic expansion of the $G$-function for the large values of the argument $z=g^{N_k}$
\begin{equation}
\fl G^{N_k,0}_{0,N_k}(z|b_1,b_2,...,b_{N_k})=\left(\frac{(2\pi)^{N_k-1}}{N_k}\right)^{1/2}e^{-N_kz^{1/N_k}}z^{\gamma}(1+K_1z^{-1/N_k} + \mathcal O(z^{-2/N_k})),
\end{equation}
where $\gamma=\frac{1-N_k}{2N_k}+\frac1N_k\sum_{n=1}^{N_k}b_n$, $K_1\sim N_k^3$. In this case the ratio in Eq.\ (\ref{answer2}) scales as $1/g\sim N_k^{-3-\varepsilon}$ and therefore contact terms vanish at large $k$
\begin{equation}
\big\langle\omega_\phi(z_1)\omega_\phi(z_2)\big\rangle_{g\sim N_k^{3+\varepsilon}}= \omega_0(z_1)\omega_0(z_2).
\end{equation}
We expect that the same result holds for the standard scaling $g\sim N_k$ \cite{FKZ}.

\section{Conclusions}

In this paper, we considered random K\"ahler metrics defined through the Bergman metric construction by random matrix measures of the eigenvalue type \eref{eig}. Metric correlators for these measures involve nontrivial angular integrals. We were able to compute several representations of the two-point functions and in particular perform the analysis of the large $k$ asymptotics. We considered in full details the particular example of the Wishart measure, where the one and two-point functions can be computed explicitly at finite $k$.

The cases we have studied clearly are oversimplified models of random metrics. For example, the form of the large $k$ two-point functions (\ref{res}) at non-coincident points is consistent with a $\delta$-function like measure, centered on the background metric $\omega_{0}$, when viewed as a measure on full space of metrics  $\kcalomega$. This singular behavior may be smoothed to some extent by considering random measures, having $\frac1k\log P$ instead of $P$ as the basic (hermitian) random matrix variable \cite{FKZ}, which we plan to do in future work. However, more interesting and physically relevant models, like the Liouville and Mabuchi theories, obviously involve a matrix model measure depending crucially on the angular part of the matrix $P$. In spite of these obvious shortcomings, we believe that our present study of the geometric matrix model correlators (\ref{integrals}) in the simplest 
``solvable'' class of models will be helpful in understanding the more physically relevant cases which are presently under investigation.

\vspace{.5cm}

\subsection*{Acknowledgements}

SK is grateful to A.~Alexandrov for useful discussions on matrix models. This work is supported in part by the belgian Fonds de la Recherche
Fondamentale Collective (grant 2.4655.07), the belgian Institut
Interuniversitaire des Sciences Nucl\'eaires (grant 4.4505.86), the
Interuniversity Attraction Poles Programme (Belgian Science Policy), the russian RFFI grant 11-01-00962 and the american NSF grant DMS-0904252.

\vspace{1cm}

\appendix
\section{Orthogonal polynomials for the Wishart ensemble}

Here we rederive (\ref{answer}) using the method of orthogonal polynomials. 

It is convenient to scale away the coupling constant $g$ in the Wishart potential, as in Eq.\ (\ref{Meijer}). Then for the linear potential $V(\lambda)=\l$ the relevant basis of orthogonal polynomials (\ref{poly}) is given by the modified Laguerre polynomials \cite{BE}, satisfying 
\begin{eqnarray}
\label{laguerre}
&&\int_0^\infty\, p_n(\lambda)p_m(\l)e^{-\l} \l^{a}d\lambda =h_n(a)\delta_{nm},\\&&\nonumber h_n(a)=\Gamma(n+1)\Gamma(n+1+a)
\end{eqnarray}
The integral here converges as long as ${\rm Re}\,a>-1$, which is true in our case. 

Rescaling $\l\rightarrow\l/g$ in Eq. (\ref{general}) and using the representation (\ref{2pforth}) we get
\begin{eqnarray}\nonumber \label{2pflag}
\fl \E_k \, \omega_{\phi_P}(z_1)\omega_{\phi_P}(z_2)=\\
\fl \nonumber=\frac1{Z_gk^2}\p\bp|_{z_1}\p\bp|_{z_2}
\lim_{t_1,t_2\rightarrow 0}|s(z_1)|^{-2t_1}|s(z_2)|^{-2t_2} \\
\fl \nonumber\cdot \int_0^\infty dx_1 \,x_1^{t_1-1} \int_0^\infty dx_2 \,x_2^{t_2-1}\frac{N_k!(N_k-1)!}{(\psi_1\psi_2)^{N_k-2}(\psi_1-\psi_2)} \int_{-i\infty}^{i\infty}da\,g^{-N_k(N_k+a)+2N_k-3} \\
\fl \nonumber \cdot\int_{\mathbb R_+^{N_k}} \Delta(\lambda)\Delta_{12}(\lambda)e^{-\lambda_1\psi_1/g- \lambda_2\psi_2/g} e^{-\sum_{i=1}^{N_k}\l_i}[d\lambda] \cdot \vol\, U(N_k)=\\
\fl \nonumber=\frac1{Z_gk^2}\p\bp|_{z_1}\p\bp|_{z_2}
\lim_{t_1,t_2\rightarrow 0}|s(z_1)|^{-2t_1}|s(z_2)|^{-2t_2} \int_0^\infty dx_1 \,x_1^{t_1-1} \int_0^\infty dx_2 \,x_2^{t_2-1}\\
\fl  \nonumber\cdot \frac{N_k!(N_k-1)!(N_k-2)!}{(\psi_1\psi_2)^{N_k-2}(\psi_1-\psi_2)} \int_{-i\infty}^{i\infty}da\,g^{-N_k(N_k+a)+2N_k-3}\prod_{n=0}^{N_k-3}h_n(a) \cdot \vol\, U(N_k)\\
\fl \nonumber \cdot \int_0^\infty d\lambda_1\,\l_1^a\int_0^\infty d\lambda_2\,\l_2^a\,\, e^{-(1+\psi_1/g)\lambda_1-(1+\psi_2/g)\lambda_2}\\
\fl \cdot(p_{N_k-2}(\lambda_1)p_{N_k-1}(\lambda_2)-p_{N_k-1}(\lambda_1)p_{N_k-2}(\lambda_2)).
\end{eqnarray}
Using the Laplace transform formula \cite{BE} for the modified Laguerre polynomials
\begin{equation}
\int_0^\infty e^{-s\l}\l^ap_n(\l)d\l=\frac{(1-s)^n}{s^{n+1+a}} \Gamma(n+1+a)
\end{equation}
we can compute the $\l$-integrals in (\ref{2pflag})
\begin{eqnarray} \nonumber
&& \int_0^\infty d\lambda_1\,\l_1^a\int_0^\infty d\lambda_2\,\l_2^a\,\, e^{-(1+\psi_1/g)\lambda_1-(1+\psi_2/g)\lambda_2}\\&&\cdot\nonumber\bigl(p_{N_k-2}(\lambda_1)p_{N_k-1}(\lambda_2)-p_{N_k-1}(\lambda_1)p_{N_k-2}(\lambda_2)\bigr)=\\\nonumber&&
=g^{-2N_k+3}\frac{h_{N_k-1}(a)h_{N_k-2}(a)}{(N_k-1)!(N_k-2)!}\cdot\frac{(\psi_1\psi_2)^{N_k-2}(\psi_1-\psi_2)}{(1+\psi_1/g)^{N_k+a}(1+\psi_2/g)^{N_k+a}}
\end{eqnarray}
Plugging this result back to  (\ref{2pflag}), and rescaling $x_{1,2}\rightarrow gx_{1,2}$ we get for the two-point function
\begin{eqnarray}
\fl \nonumber\E_k \, \omega_{\phi_P}(z_1)\omega_{\phi_P}(z_2)=\\\nonumber
\fl = \frac1{Z_gk^2}\p\bp|_{z_1}\p\bp|_{z_2}
\lim_{t_1,t_2\rightarrow 0}|s(z_1)|^{-2t_1}g^{t_1}|s(z_2)|^{-2t_2} g^{t_2} \int_{-i\infty}^{i\infty}da\,g^{-N_k(N_k+a)}\\\nonumber
\fl \cdot\int_0^\infty dx_1 \,x_1^{t_1-1} \int_0^\infty dx_2 \,x_2^{t_2-1}\bigl[(1+\psi_1)(1+\psi_2)\bigr]^{-(N_k+a)} N_k!\prod_{n=0}^{N_k-1}h_n(a) \cdot \vol\, U(N_k)=
\\\fl \nonumber= \int_{-i\infty}^{i\infty}da \frac1{k^2}\frac{Z_{g,a}}{Z_g}\p\bp|_{z_1}\p\bp|_{z_2}
\lim_{t_1,t_2\rightarrow 0}|s(z_1)|^{-2t_1}g^{t_1}|s(z_2)|^{-2t_2} g^{t_2} \\
\fl  \nonumber
\cdot\int_0^\infty dx_1\,x_1^{t_1-1}\int_0^{\infty}dx_2 \,x_2^{t_2-1}(1+x_1+x_2+ x_1x_2\sin^2\theta_{12})^{-(N_k+a)}\\
\fl  \nonumber = \int_{-i\infty}^{i\infty}da\, \E^a_k \, \omega_{\phi_P}(z_1)\omega_{\phi_P}(z_2),
\end{eqnarray}
which coincides precisely with (\ref{2pointf1}).

\end{document}